# Inhomogeneous Energy-Density Driven Instability in Presence of a Transverse DC Electric Fields in a Magnetized Plasma Cylinder

**Suresh C. Sharma**[*] **and Satoshi Hamaguchi**
Science and Technology Center for Atoms, Molecules, and Ions Control, Graduate School of Engineering, Osaka University, 2-1 Yamada-oka, Suita, Osaka 565-0871, Japan
*Currently on Leave from the Physics Department, GPMCE (G.G.S. Indraprastha University,
 Delhi), **INDIA**

**ABSTRACT**
   Temporal evolution of the inhomogeneous energy density driven instability (IEDDI) is examined in presence of a nonuniform transverse dc electric field in collisional magnetized plasma cylinder. Using experimentally known parameters relevant to IEDDI, we have estimated the growth rate for $\nu$=1, 2 and 3 nonlocal eigenmodes.

## 1. INTRODUCTION

   IEDD waves propagate primarily in the $\vec{E} \times \vec{B}$ direction and have frequencies in a wide range near the ion cyclotron frequency. For $R(r) = k_\perp v_E(r)/k_z v_{de} \ll 1$, the case of large $v_{de}$ and small $v_E$, the electrostatic ion cyclotron (EIC) waves are current-driven (EIC Instability)[1,2]. For $R \gg 1$, the case of small $v_{de}$ and large $v_E$, the EIC waves are driven by the transverse, localized, dc electric field (IEDD Instability)[3-4].

## 2. Instability Analysis

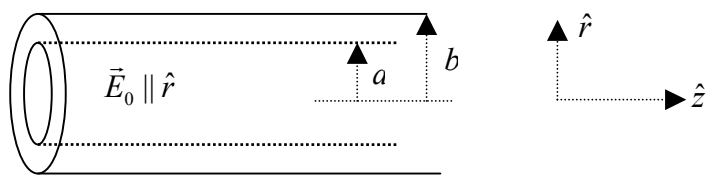

Fig. 1 A schematic diagram of the electric
       field model

**P**lasma waveguide radii $b_1$, $a_1$, equilibrium density $n_{op}$, $T_e \approx T_i$, collisional frequency (= $\nu_e$ for electrons and =0 for ions), static magnetic field $B$ is in the $z$-direction, dc electric field $E_0(r)$ is in the radial direction, drift $\vec{v}(r) = \dfrac{c\vec{E}_0(r)\times\vec{B}}{B^2}$, Electric fields $E_0(r)$ model

$E_0(r) = E_0$    for $0<r<a_1$    I region
        $= 0$    for $a_1<r<b_1$    II region

In addition electrons have a flow or drift ($v_{de}$) along the magnetic field direction. A nonlocal wave packet can couple these two regions, and a flow of energy from region I to region II enables the IEDD mode to grow. This gives rise to IEDDI.

**Basic Equations**

$$m_\alpha \frac{d\vec{v}_\alpha}{dt} = q_\alpha \vec{E} - \frac{e}{c}\vec{v}_\alpha \times \vec{B} - \frac{\nabla(p_\alpha)}{n_\alpha} - m_\alpha \nu_e \vec{v}_\alpha. \tag{1}$$

$$\frac{\partial n_\alpha}{\partial t} + \nabla \cdot (n_\alpha \vec{v}_\alpha) = 0 \tag{2}$$

$$\nabla \cdot \vec{E} = 4\pi \rho_\alpha \tag{3}$$

where $\alpha = e, i$.

By solving equations (1)-(3), we obtain $\nu$th order Bessel equation for $\Phi_1$

$$\frac{\partial^2 \Phi_1}{\partial \eta^2} + \frac{1}{\eta}\frac{\partial \Phi_1}{\partial \eta} + (1 - \frac{\nu^2}{\eta^2})\Phi_1 = 0, \tag{4}$$

$\eta = k_I r$      for $0<r<a_1$,
     $= k_{II} r$      for $a_1<r<b_1$,

$$k_I^2 = \dfrac{[\dfrac{\omega_p^2}{v_{te}^2 k_z^2(-1+\dfrac{i\nu_e \omega_1}{k_z^2 v_{te}^2})} - 1]k_z^2}{[1+\dfrac{\omega_p^2}{\omega_c^2} - \dfrac{\omega_{pi}^2}{\omega_2^2 - \omega_{ci}^2}]}. \tag{5}$$

$$k_{II}^2 = \frac{[\frac{\omega_p^2}{v_{te}^2 k_z^2(-1+\frac{i\nu_e(\omega-k_z v_{de})}{k_z^2 v_{te}^2})}-1]k_z^2}{[1+\frac{\omega_p^2}{\omega_c^2}-\frac{\omega_{pi}^2}{\omega^2-\omega_{ci}^2}]}. \qquad (6)$$

Now dividing Eq. (5) by (6), and cross-multiplying, we get

$$\epsilon_r(\omega,k) + i\,\epsilon_i(\omega,k) = 0, \qquad (7)$$

where

$$\varepsilon_r(\omega,k) = 1 + \frac{\omega_p^2}{k_z^2 v_{te}^2} - \frac{\omega_p^2}{k_z^2 v_{te}^2}\frac{\omega_{pi}^2}{[(\omega_2^2-\omega_{ci}^2)]}\frac{k_I^2}{(k_I^2-k_{II}^2)(1+\frac{\omega_p^2}{\omega_c^2})}$$
$$+\frac{\omega_p^2}{k^2 v_{te}^2}\frac{\omega_{pi}^2}{(\omega^2-\omega_{ci}^2)}\frac{k_I^2}{(k_I^2-k_{II}^2)(1+\frac{\omega_p^2}{\omega_c^2})} -$$
$$\frac{\omega_{pi}^2}{[(\omega_2^2-\omega_{ci}^2)]}\frac{k_I^2}{(k_I^2-k_{II}^2)(1+\frac{\omega_p^2}{\omega_c^2})} - \frac{\omega_{pi}^2}{(\omega^2-\omega_{ci}^2)}\frac{k_{II}^2}{(k^2-k_{II}^2)(1+\frac{\omega_p^2}{\omega_c^2})},$$

(8)

$$\varepsilon_i(\omega,k) = \nu_e[\frac{\omega_1}{k_z^4 v_{te}^4}\frac{\omega_p^2 \omega_{pi}^2}{(\omega^2-\omega_{ci}^2)}\frac{k_I^2}{(k_I^2-k_{II}^2)(1+\frac{\omega_p^2}{\omega_c^2})} - \frac{\omega_1}{k_z^4 v_{te}^4}\frac{\omega_p^2 \omega_{pi}^2}{(\omega^2-\omega_{ci}^2)} \times$$
$$\frac{k_{II}^2}{(k_I^2-k_{II}^2)} + \frac{(\omega-k_z v_{de})}{k_z^4 v_{te}^4}\frac{k_I^2 \omega_p^2}{(k_I^2-k_{II}^2)} - \frac{(\omega-k_z v_{de})}{k_z^4 v_{te}^4}\frac{k_I^2 \omega_p^2}{(k_I^2-k_{II}^2)} \times$$
$$\frac{\omega_{pi}^2}{[(\omega_2^2-\omega_{ci}^2)](1+\frac{\omega_p^2}{\omega_c^2})}].$$

(9)

We write $\omega = \omega_r + i\gamma$ ($|\gamma| \ll \omega_r$). The real and imaginary part of the frequencies are given by

$$\omega_r = k_\theta v_E + [\omega_{ci}^2 + \frac{k_I^2}{k_p^2(k_I^2-k_{II}^2)} \frac{k_z^2 c_s^2}{(1+\frac{\omega_p^2}{\omega_c^2})}]^{1/2}. \tag{10}$$

$$\omega_r = k_\theta v_E + [\frac{\omega_{ci}^2 \frac{k_I^2}{(k_I^2-k_{II}^2)} \frac{k_z^2 c_s^2}{(1+\frac{\omega_p^2}{\omega_c^2})k_p^2}}{\omega_{ci}^2 + \frac{k_I^2}{(k_I^2-k_{II}^2)} \frac{k_z^2 c_s^2}{(1+\frac{\omega_p^2}{\omega_c^2})k_p^2}}]^{1/2}. \tag{11}$$

$$\gamma = v_e [\frac{\{1-\frac{k_z v_{de}(1+R)}{\omega_r}\}\omega_p^2 \omega_{pi}^2 k_I^2 \omega_p^2}{k_z^4 v_{te}^4 \omega_r (k_I^2-k_{II}^2)\omega_c^2(1+\frac{\omega_p^2}{\omega_c^2})} - \frac{(\omega_r-k_z v_{de})k_I^2 \omega_p^2}{k_z^4 v_{te}^4 (k_I^2-k_{II}^2)}(1-\frac{\omega_{pi}^2}{[(\omega_r-k_\theta v_E)^2-\omega_{ci}^2](1+\frac{\omega_p^2}{\omega_c^2})})] /$$

$$2\frac{\omega_{pi}^4}{k_z^2 c_s^2} \frac{1}{(k_I^2-k_{II}^2)(1+\frac{\omega_p^2}{\omega_c^2})} [\frac{(\omega_r-k_\theta v_E)k_I^2}{[(\omega_r-k_\theta v_E)^2-\omega_{ci}^2]^2} + \frac{k_{II}^2}{\omega_r^3}],$$

(12)

Equation (10) gives the real frequency of IEDD modes in presence of a transverse dc electric fields.

**RESULTS**
### 3. Parameters[3] :
1. Finite gyroradius parameter $b(=k_\theta^2 \rho_i^2)$ =0.1 -0.8, $n_{p0} \approx 1 \times 10^9$ cm$^{-3}$ , $T_e$ = $T_i \approx 0.2$ eV , $m_i/m$ (Sodium) = $4.6 \times 10^4$, $k_\square = 2.8$ cm$^{-1}$, $k_z = 0.65$ cm$^{-1}$, $v_e = 3.22 \times 10^8$ sec$^{-1}$, $v_E = 3.3 \times 10^8$/B(in Gauss), $v_{de} = $ ~$1.76 \times 10^4$ cm/sec.

We have plotted in Fig. 2 the normalized growth rate ($\gamma/\omega_{ci}$) [ a) for $\nu$=1, b) for $\nu$=2 and c) for $\nu$=3 eigenmodes] of the IEDD instability as a function of the finite gyroradius parameter $b$ for the same parameters as mentioned above. The growth rate of the instability increases with the finite gyroradius parameter $b$ and reaches maximum in accordance with the experimental observations by Koepke et al[3]

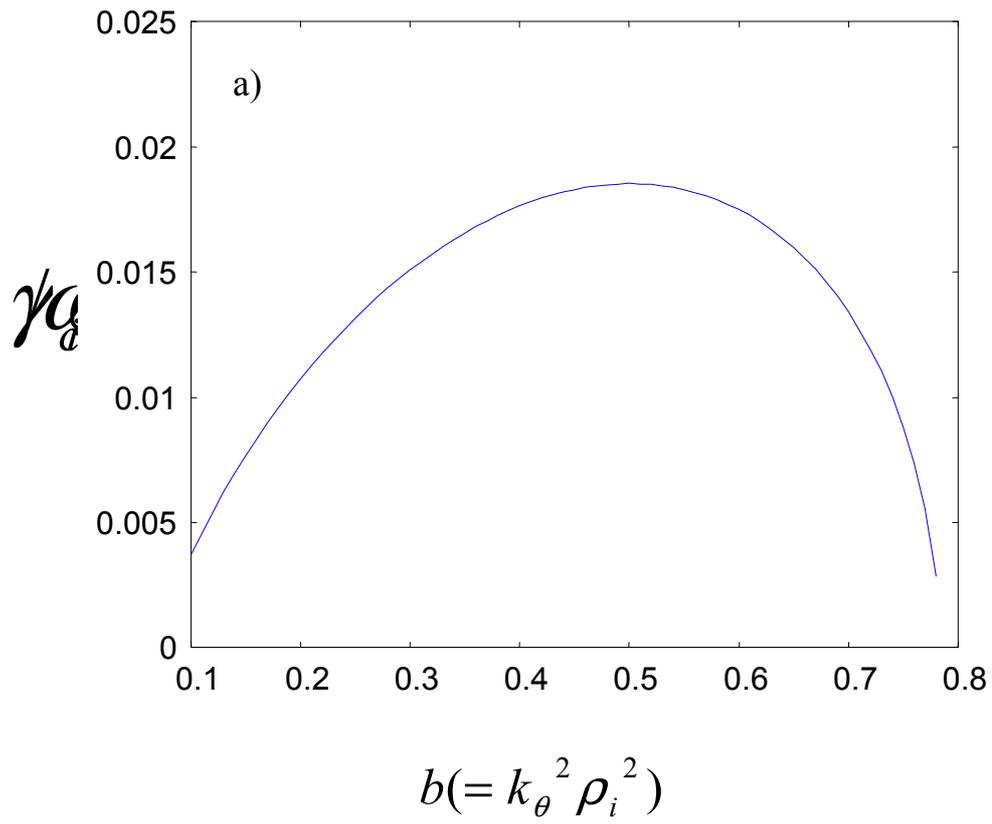

a)

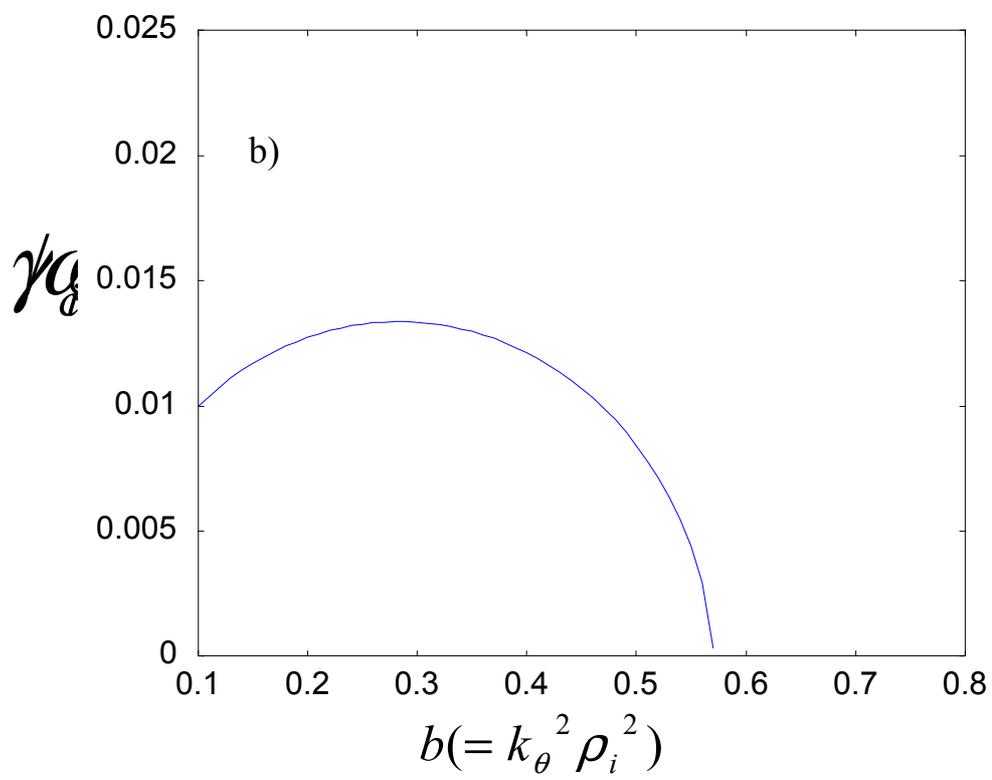

b)

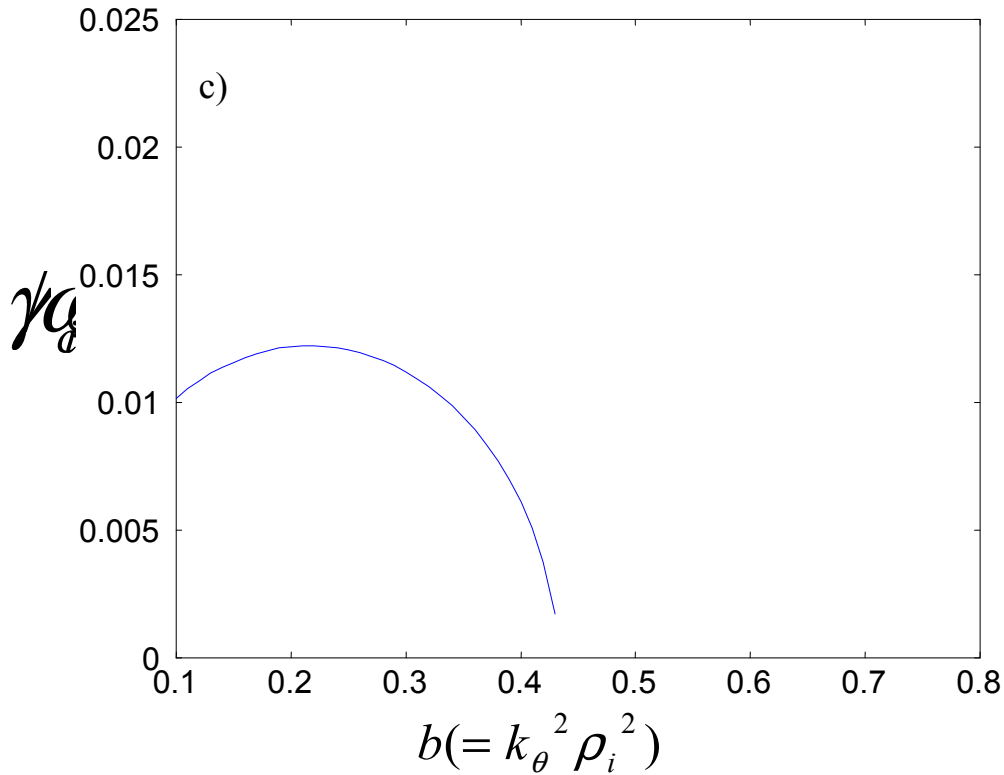

Fig. 2 : Normalized growth rate ( $\gamma/\omega_{ci}$ ) of the inhomogeneous energy-density driven instability as a function of the finite gyroradius parameter $b(=k_\theta^2 \rho_i^2)$ [ a) for ν=1 b) for ν=2, and c) for ν=3 eigenmodes]. The parameters are given in the text.